# Link Reassignment based Snapshot Partition for Polar-orbit LEO Satellite Networks

Zhu Tang, Zhenqian Feng, WanrongYu, Baokang Zhao, Chunqing Wu
College of Computer, National University of Defense Technology, Changsha, China
Email: {tangzhu, wlyu, bkzhao, wuchunqing}@nudt.edu.cn, z_q_feng@163.com

*Abstract*—Snapshot is a fundamental notion proposed for routing in mobile low earth orbit (LEO) satellite networks which is characterized with predictable topology dynamics. Its distribution has a great impact on the routing performance and on-board storage. Originally, the snapshot distribution is invariable by using the static snapshot partition method based on the mechanical steering antenna. Utilizing nowadays advanced phased-array antenna technology, we proposed a quasi-dynamic snapshot partition method based on the inter-satellite links (ISLs) reassignment for the polar-orbit LEO satellite networks. By steering the inter-satellite antennas when snapshot switches, we can reassign the ISLs and add available ISLs for a better following snapshot. Results show that our method can gain more stable routing performance by obtaining constant snapshot duration, constant ISL number and lower end to end delay. Especially for Iridium system, our method can gain not only longer merged snapshot duration, half of the snapshot number, but also higher utilization ratio of inter-plane ISLs. Potentially, our method could be very useful for the Iridium-next system.

## I. INTRODUCTION

Compared to the GEO and MEO satellites, LEO satellites have shorter transmission latency and lower signal attenuation, which is more desirable for the voice communication and real-time data interaction. However, the rapid moving and internetworking property of LEO satellites lead to frequent handoff on the inter-satellite links and up-down links, making the topology of satellite network change acutely.

With the industrial development of satellite network, a lot of routing algorithms have been proposed for LEO satellite networks, which can be divided into static routing algorithms such as virtual topology[1][2][3][4], virtual node[5][6], and dynamic routing algorithms such as on-demand routing algorithm[7], traffic-aware routing algorithm[8] and so on. The dynamic routing algorithms which are inspired by the terrestrial wired and ad-hoc routing algorithms, are more flexible and can react rapidly to traffic changes and link failures. However, these complex functions will consume a lot of computation and bandwidth resources which are limited in satellite networks. On the other hand, the static routing algorithms take full advantages of the periodicity and predictability of LEO satellite networks to weaken the impact caused by the dynamic topology, avoiding large onboard calculation and distributed communication. In fact, current operational polar-orbit LEO satellite network - Iridium system employs the virtual topology based snapshot routing algorithm[9]. Therefore, the static routing algorithm is more suitable for LEO satellite networks, and it is very meaningful to improve the snapshot routing algorithm's performance potentially for the Iridium-next system.

The main idea of the snapshot routing algorithm recently formulated in [4] is to cut the orbit period into many small time slices which contain temporary stable network topology, i.e. the connectivity of ISLs is unchanged in each time slice. Due to the predictability and periodicity of satellite movements, the routing table for each snapshot can be pre-calculated and stored on-board for switching according to the precise timetable. In this way, the constellation topology dynamics are concealed by many successive short-term stable snapshots. Consequently, the distribution of snapshots has great effects on the routing performance of satellite networks. Improving the quality of the snapshot distribution will certainly improve the routing performance. Currently there are two static partition methods to obtain the snapshot distribution:

- Fixed ISL on-off partition: Gounder[1] proposed that once an ISL in the satellite network is established or broken down, a new snapshot is produced. The theoretical analysis and simulation validation of this method is presented in [10].

- Equal time interval partition: Werner[2] proposed to use equal time interval for snapshot partition, because the shortest possible time slice can properly adapt to the link latency variation. Wang J. L. et al.[11] improved this method to prolong the snapshot duration by eliminating ISLs whose connectivity will change in the longer time interval.

Due to the high frequency of ISL on-offs in polar orbit constellation, the ISL on-off partition method will create many short duration snapshots, and the frequent routing table switches will affect the routing stability of satellite networks. The equal time interval partition method can efficiently reduce the packet loss rate and on-board storage requirement, but the latency variation of ISLs cannot be revealed if the duration becomes too long. Meanwhile, the ISLs utility is poor if many non-persistent ISLs are eliminated.

The above unavoidable tradeoff roots in the inflexibility of existing static link assignment method. As the early on-board inter-satellite links employ the mechanical steering antennas, they could not move fast enough to create a brand new ISL with another peer antenna. At present, the phased-array antenna with electronic steering ability makes the ISL's rearrangement possible and easy. Furthermore, BEIDOU II navigation system intends to apply this advanced technology in

its ISLs for fast navigation information exchange. To this end, we'd like to ask a question:

*Is it possible and how to optimize the snapshot distribution in polar-orbit LEO satellite networks through link reassignment?*

Previous studies have utilized the dynamic link reassignment in LEO satellite networks for adaptively adjusting the network performance[12][13] and maximizing the minimum residual capacity links[3]. However, the snapshot partition problem is not considered yet. In this paper, we propose an link reassignment based snapshot partition method in polar-orbit LEO satellite networks. When current snapshot ends, all the ISLs are reassigned with universal techniques for a better subsequent snapshot quality. The validation simulations are performed on two typical polar-orbit LEO satellite networks: Iridium system and Teledesic system. The results show that our method owns important improvements on the snapshot distribution, utilization ratio of inter-plane ISLs and end to end delay.

The rest of the paper is organized as follows. In Section 2, we present the basic polar-orbit LEO satellite network model and the feasibility of phased-array antennas applying to satellite networks. In Section 3, we propose the link reassignment based snapshot partition method for polar-orbit LEO satellite networks. The proposed method is analyzed and validated by simulations in Section 4. Finally, we summarize the paper in Section 5.

## II. BACKGROUND

### A. Polar-orbit Satellite Network Model

Assuming that there are N orbit planes in the polar-orbit satellite network, and each plane has M satellites uniformly distributed, so the total number of satellites in the constellation is N*M. Satellites in the same plane are separated by angle $\omega = \frac{2\pi}{M}$. Each planes are separated in range of $\pi$ by the same angle $\Delta\Omega$. To provide the global coverage with minimized overlapping, the phase angle of satellites in adjacent planes is $\omega_f = \frac{\omega}{2} = \frac{\pi}{M}$, which is shown in Fig. 1. Let satellite $S_{ij}$ denote the jth satellite in plane i, i=1,...,N, j=1,...,M. Let $i \in (0, \pi)$ represent the inclination of the orbit plane, h represent the orbit altitude above the ground and T represent the orbit period of satellites. The terms and reference parameter values of Iridium and Teledesic systems are summarized in TABLE I.

TABLE I. TERMS AND REFERENCE VALUES IN SATELLITE SYSTEM

| Parameters | Iridium | Teledesic |
| --- | --- | --- |
| Number of planes (N) | 6 | 12 |
| Satellites per plane (M) | 11 | 24 |
| Inclination (i) | 86.4° | 84.7° |
| Altitude (h) | 780km | 1375km |
| The orbit period of the satellite network (T) | 100.45 | 113.23 |
| Angle between planes with same moving direction ($\Delta\Omega$) | 31.6° | 15.36° |
| Phase angle of satellites in adjacent planes ($\omega_f$) | 16.36° | 7.5° |

As is shown in Fig. 1, polar orbit constellation contains two kinds of ISLs:

- Intra-plane ISLs: they connect the up and down adjacent satellites in the same plane, such as ISL ($S_{i,j}$, $S_{i,j+1}$) and ($S_{i,j}$, $S_{i,j-1}$). The relative position of satellites in the same plane keeps unchanged, so the intra-plane ISLs maintain permanently.

- Inter-plane ISLs: they connect the left and right satellites in adjacent orbits, such as ($S_{i,j+1}$, $S_{i+1,j+1}$) and ($S_{i+2,j+1}$, $S_{i+1,j+1}$). The connectivity and latency of these ISLs change with the satellite movements. But because the adjacent satellites in the first and last plane have opposite moving directions, there is no ISLs established between them. Therefore, the satellites near the seam are equipped with only one inter-plane ISLs, while the others contain two.

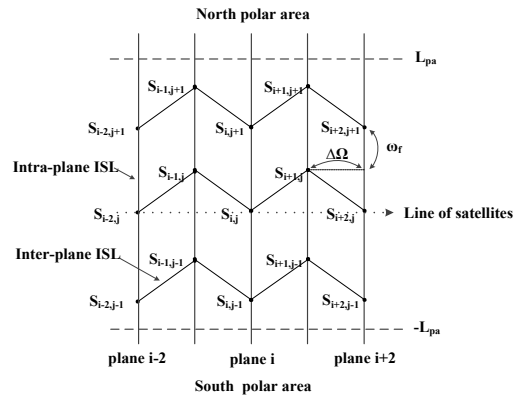

Fig. 1. Polar-orbit satellite network topology

Because the satellites in adjacent planes will exchange positions from left to right, the relative speed between satellites near the polar will exceed the automatic tracking speed of current on-board antennas. So the polar-orbit constellation defines the north and south polar area for closing the inter-plane ISLs. Until satellites exit from the polar areas from the other hemisphere of the earth, the inter-plane ISLs can be established again. The latitudes of the north and south polar area border are usually equivalent and denoted by $L_{pa}$ and -$L_{pa}$, respectively. For example, when $S_{i+1,j+1}$ enters the north polar area, the inter-plane ISLs ($S_{i,j+1}$, $S_{i+1,j+1}$) and ($S_{i+2,j+1}$, $S_{i+1,j+1}$) should be switched off. Until the satellites $S_{i+1,j-1}$, $S_{i,j-1}$ and $S_{i+2,j-1}$ all have exited from the polar area, the inter-plane ISLs can be established again.

### B. Feasibility Analysis of Phased-Array Antenna

The S-band phased-array antenna for inter-satellite link has been tested in Engineering Test Satellite VI (ETS-VI) in 1995[14]. Recently, many high frequency phased-array antennas are designed for ISLs. Cowley et al.[15] designed a Ka-band sectorised and a V-band torus phased-array antenna for the ISLs of LEO satellite networks. Results show that the gains and EIRPs of the new antennas are quite sufficient for the LEO ISLs.

Practically, phased-array antenna's electronic steering property can easily achieve large elevation (0°~90°) and azimuth (0º~360º) than traditional mechanical steering antenna. Larger antenna gain for longer distance communication when steering can be obtained by activating more feed elements in the phased-array[15]. In sum, the phased-array antenna technology is quite mature for applying in ISLs of satellite networks. Due to its fast steering feature, the phased-array antenna based link reassignment method can be feasible in LEO satellite networks under current basic antenna constraints.

### III. LINK REASSIGNMENT BASED SNAPSHOT PARTITION

Regular connectivity changes of inter-plane ISLs lead to the regularity of snapshot distribution in polar-orbit constellation. For polar-orbit LEO satellite networks, we define below two local inter-plane ISLs for better snapshot routing performance, which are defined as follows:

**Definition 1** Oblique inter-plane ISLs ($link_{oblique}$): As is shown in Fig. 2, the inter-plane ISLs between satellites in adjacent planes, such as the inter-plane ISLs between $S_{i,j}$ and $S_{i+1,j}$, $S_{i+1,j-1}$, $S_{i-1,j}$, $S_{i-1,j+1}$, are called oblique inter-plane ISLs.

**Definition 2** Horizontal inter-plane ISLs ($link_{horizontal}$): As is shown in Fig. 2, the inter-plane ISLs between satellites separated by a plane and owning the same latitude and same moving direction, such as links between $S_{i,j}$ and $S_{i-2,j}$, $S_{i+2,j}$, are called horizontal inter-plane ISLs.

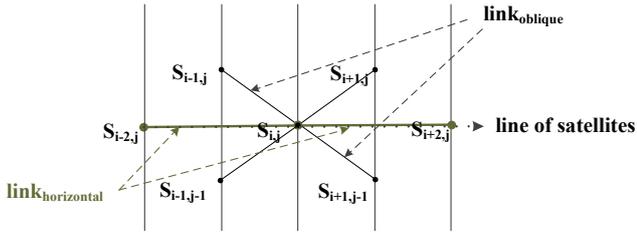

Fig. 2. Oblique and horizontal inter-satellite links and line of satellites

In most polar-orbit satellite systems, the oblique inter-plane ISLs are full connected in the whole orbit period. Since the horizontal inter-plane ISLs can be blocked by earth and atmosphere, the survival latitude ranges of horizontal inter-plane ISLs are defined such as ($L_{horizontal}$, $L_{pa}$) for the north hemisphere and ($-L_{horizontal}$, $-L_{pa}$) for the south hemisphere. $L_{horizontal}$ can be calculated as follows,

$$L_{horizontal} = arcsin\left(\sqrt{\frac{cos(\theta_{max})-cos(2*\Delta\Omega)}{1-cos(2*\Delta\Omega)}}\right) \quad (1)$$

where $\theta_{max}$ is the maximal geocentric angle between two attachable satellites, and the $\Delta\Omega$ is the angle between adjacent orbit planes. For the Iridium system, $L_{horizontal} = 32.81º$, which is much smaller than the polar border range as $60º \sim 75º$. Furthermore, $link_{horizontal}$ is more suitable for antenna alignment on account of its lower relative angular velocity between satellites.

Therefore, when satellites come close to the polar area, we establish the horizontal inter-plane ISLs, such as ($S_{i,j}$, $S_{i+2,j}$) and ($S_{i,j}$, $S_{i-2,j}$) to prolong the snapshot duration. When satellites exit from the polar area, it is better to assign oblique inter-plane ISLs for longer ISL lifetime, such as ($S_{i,j}$, $S_{i-1,j}$), ($S_{i,j}$, $S_{i-1,j-1}$), ($S_{i,j}$, $S_{i+1,j}$) and ($S_{i,j}$, $S_{i+1,j-1}$).

#### A. Method Describption

To better describe the link reassignment method, we define the concept of line of satellites in polar orbit constellation as follows:

**Definition 3** Line of satellites (LS): In polar orbit constellation, the set of satellites with the same latitude and same moving direction is called a line of satellites, which is illustrated in Fig. 2.

Let $LS_i$ denote the ith LS counting form the south pole in direction of ascending movement, i=1…NLS, and NLS represent the number of LS in the whole constellation. Let $NLS_{pa}$ denote the number of LS in one polar area (North or South polar area), and $NLS_{npa}$ denote the number of LS in one non-polar area in one hemisphere (ascending or descending hemisphere). The LSs just now exit from the both polar areas are all called $LS_{npa0}$, which is shown in Fig. 3~Fig. 5. Based on the symmetrical characteristic of polar-orbit satellite networks, NLS can be calculated as follows,

$$NLS = 2 * (NLS_{npa} + NLS_{pa}) = 2 * M \quad (2)$$

where $NLS_{npa}$ can be calculated by the phase angle of satellites in adjacent planes $\omega_f$ and the polar border latitude $L_{pa}$,

$$NLS_{npa} = \left\lfloor \frac{2L_{pa}}{\omega_f} \right\rfloor \quad (3)$$

At any given time t in the orbit period, the link reassignment can be executed normally from $LS_{npa0}$, i.e. the LS just leaves the polar area in both hemispheres, to the other border of the polar area. Single snapshot partition event (ISL break down event or ISL establishment event) is applied to gain better snapshot distribution, which is prior to existing fixed ISL on-off based partition:

*1) ISL break down event based partition*

The ISL break down event based partition is triggered by each LS entering the polar area (see the yellow satellites and ISLs in Fig. 3(a)~(d)). Based on the parity of $NLS_{npa}$ and the plane numbers on $LS_{npa0}$, there are four kinds of reassignment techniques based on the ISL break down event:

**Reassignment A1**

If $NLS_{nps}$ is even, we assign the oblique ISLs from $LS_{npa0}$ to the last LS in non-polar area. And if the satellites in $LS_{npa0}$ are in odd planes, the network topology after link reassignment is illustrated in Fig. 3(a). The number of horizontal ISLs is 0.

**Reassignment A2**

If $NLS_{nps}$ is even and the satellites in $LS_{npa0}$ are in even planes, the new network topology illustrated in Fig. 3(b) is obtained with the same ISLs assignment method as A1. Since the plane numbers of satellites in $LS_{npa0}$ will change their parity alternately for satellite's orbiting movements, Reassignments A1 and A2 are alternately applied in the specific satellite networks.

**Reassignment A3**

If $NLS_{nps}$ is odd, the oblique ISLs are assigned for the even number of LSs as above, till the last LS in non-polar area. The horizontal ISLs are assigned in this LS. And If the satellites in $LS_{npa0}$ are in odd planes, the new network topology is illustrated in Fig. 3(c).

**Reassignment A4**

If $NLS_{nps}$ is odd and the satellites in $LS_{npa0}$ are in even planes, as is illustrated in Fig. 3(d), the horizontal ISLs assignment simply seems like to be shifted from Reassignment A3. Similar to A1 and A2, Reassignments A3 and A4 are applied alternately due to the satellite movements.

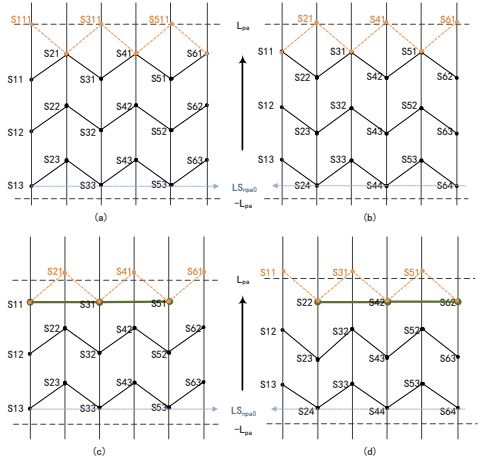

Fig. 3. Link reassignment techniques based on ISL break down event (asceding hemisphere)

*2) ISL establishment event based partition*

The ISL establishment event based partition is triggered by each LS exiting from the polar area (see the yellow satellites and links in Fig. 4 and Fig. 5). According to the number and distribution of LS in non-polar area, and the parity of the plane number in $LS_{npa0}$, there are eight kinds of link reassignments:

**Reassignment B1~B4**

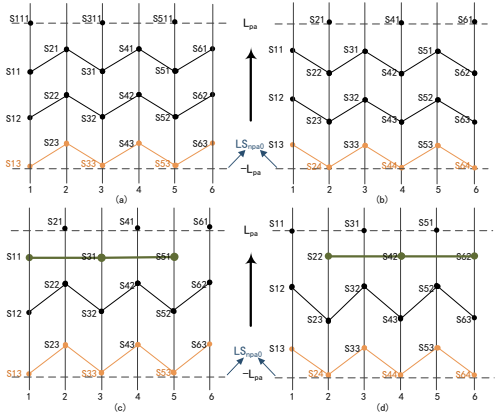

Fig. 4. Link reassignment techniques B1~B4 based on ISL establishment event (asceding hemisphere)

If $\frac{2L_{pa}}{\omega_f} \in \mathbb{N}$, the LSs are uniformly distributed in non-polar areas, i.e. when one LS enters polar area, there must be one LS exits from the polar area events. All the LSs in non-polar areas can be assigned with ISLs from the $LS_{npa0}$ to the last LS with oblique ISLs. As described in Reassignments A1~A4, the parity of the number of LS in non-polar area determines whether to assign the horizontal ISLs. Meanwhile, the parity of plane numbers in $LS_{npa0}$ also make the link assignment different. Therefore, the link reassignment techniques can be depicted in Fig. 4(a)~(d), each is corresponded to the one in Reassignments A1~A4 orderly.

**Reassignment B5~B8**

If $\frac{2L_{pa}}{\omega_f} \notin \mathbb{N}$, the LSs are nonuniformly distributed in non-polar area. The events of LS entering and exiting from the polar area are occurred alternately. To avoid the impact of link break down event to snapshot partition, the last LS in non-polar area which is about to enter the polar area is not assigned any ISLs. Based on the parity of the $NLS_{npa}$ and the plane numbers in $LS_{npa0}$, the rest of ISLs are assigned in the same way as Reassignments A1~A4, which are depicted in Fig. 5(a)~(d).

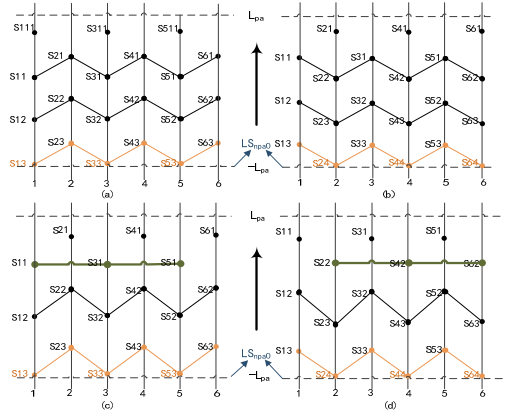

Fig. 5. Link reassignment techniques B5~B8 based on ISL establishment event (asceding hemisphere)

Fig. 3~Fig. 5 provide 12 kinds of link reassignment techniques, but for the satellite system with given constellation parameters and polar border latitude, just one pair of the reassignment techniques is enough for use fixedly all the time.

*B. Common Property*

Although based on different snapshot partition event, the two snapshot partition methods have the same properties in the aspect of snapshot duration, snapshot number and the inter-plane ISL number. The regularities of the new partition methods are concluded as follows:

**Property 1**. By using the link reassignment techniques, the snapshot duration $\delta_{dynamic}$ and snapshot number $S_{reassign}$ in polar-orbit LEO satellite networks is constant and irrelevant to the setting of the polar border latitude.

Analysis: Because the link reassignment techniques employ the single snapshot generation event for snapshot partition, the new snapshot duration $\delta_{reassign}$ is constant and determined by the behavior that LSs enter/exit from the polar area. Therefore, the snapshot duration equals to the time interval of successive such two events. Given the number of LS in polar orbit constellation as NLS, the snapshot duration $\delta_{reassign}$ can be calculated as follows,

TABLE II. SNAPSHOT DISTRIBUTIONS OF SATELLITE SYSTEMS BY THEORETICAL CALCULATION AND SIMULATIONS

| System | $L_{pa}$ | $\delta_{reassign}/\widehat{\delta_{reassign}}$ | $NISL_{reassign}/\widehat{NISL_{reassign}}$ | $S_{reassign}/\widehat{S_{reassign}}$ | $\widehat{\delta_{fixed}}$ max | $\widehat{\delta_{fixed}}$ min | $\widehat{NISL_{fixed}}$ max | $\widehat{NISL_{fixed}}$ min | $\widehat{S_{fixed}}$ | $\widehat{\delta_{equal}}$ | $\widehat{NISL_{equal}}$ | $\widehat{S_{equal}}$ |
|---|---|---|---|---|---|---|---|---|---|---|---|---|
| Iridium (6*11) | 60° | 274.31/274.31 | 34/34 | 22/22 | 179.90 | 92.10 | 30 | 35 | 44 | 274.31 | 30(25) | 22 |
|  | 65° | 274.31/274.31 | 34/34 | 22/22 | 261.91 | 10.11 | 35 | 30 | 44 | 274.31 | 30(25) | 22 |
|  | 70° | 274.31/274.31 | 40/40 | 22/22 | 158.29 | 113.75 | 40 | 35 | 44 | 274.31 | 35(30) | 22 |
|  | 75° | 274.31/274.31 | 44/44 | 22/22 | 215.75 | 56.33 | 40 | 45 | 44 | 274.31 | 35(40) | 22 |
| Teledesic (12*24) | 60° | 141.69/141.69 | 176/176 | 48/48 | 157.73 | 125.35 | 176 | 154 | 48 | 141.69 | 154(176) | 48 |
|  | 65° | 141.69/141.69 | 186/186 | 48/48 | 215.78 | 67.39 | 176 | 198 | 48 | 141.69 | 176(154) | 48 |
|  | 70° | 141.69/141.69 | 198/198 | 48/48 | 261.83 | 21.24 | 198 | 176 | 48 | 141.69 | 176(198) | 48 |
|  | 75° | 141.69/141.69 | 220/220 | 48/48 | 177.30 | 105.78 | 220 | 198 | 48 | 141.69 | 198(220) | 48 |

$$\delta_{reassign} = \frac{T}{NLS} = \frac{T}{2*M} \quad (4)$$

Because the snapshot duration keeps constant, the number of snapshots in polar orbit constellation $S_{reassign}$ also keeps constant,

$$S_{reassign} = \frac{T}{\delta_{reassign}} = 2*M \quad (5)$$

**Property 2**. If the constellation parameters and the polar border latitude are given, the number of inter-plane ISLs denoted by $NISL_{reassign}$ is determined and keeps constant in all snapshots.

Analysis: Due to the symmetrical distribution of satellites in polar-orbit constellation, every snapshot has only one pair of symmetrical LSs entering/exiting from the north and south polar area. Given the constellation parameters and polar border latitude, the number and location of LS in the non-polar area keep stable in each snapshot, and the link reassignment employs just one pair of the topologies illustrated in Fig. 3~Fig. 5. Therefore, the number of inter-plane ISLs in each snapshot ($NISL_{reassign}$) is constant, which can be calculated as follows,

$$NISL_{reassign} = NISL_{oblique} + NISL_{horizontal} \quad (6)$$

where the number of oblique ISLs ($NISL_{oblique}$) and horizontal ISLs ($NISL_{horizontal}$) can be calculated as follows,

$$NISL_{oblique} = 2 * \left\lfloor \frac{NLS_{npa}}{2} \right\rfloor * (N-1) \quad (7)$$

$$NISL_{horizontal} = \begin{cases} N-2, & if\ NLS_{npa} \in 2*\mathbb{N}+1 \\ 0, & if\ NLS_{npa} \in 2*\mathbb{N} \end{cases} \quad (8)$$

If the number of LS in non-polar area ($NLS_{npa}$) is odd, the inter-plane ISLs contain both the oblique ISLs and horizontal ISLs. While if the $NLS_{npa}$ is even, the inter-plane ISLs only contain the oblique ISLs.

## IV. EVALUATION

In this paper, the snapshot distribution and resulting network performance are evaluated for the link reassignment method, fixed ISL on-off partition method and the equal time partition method in network simulator NS2. All of the visibility data of Iridium and Teledesic system are obtained through the satellite orbit simulator Satellite Tool Kit (STK). The constellation parameters of Iridium and Teledesic system are given in TABLE I. .

TABLE II. shows the snapshot distribution of Iridium and Teledesic system under different polar border configurations ($L_{pa}$ = 60°, 65°, 70° and 75°). Columns $\delta_{reassign}$, $S_{reassign}$ and $NISL_{reassign}$ are theoretical results from equation (4)~(8), while $\widehat{\delta_{reassign}}$, $\widehat{S_{reassign}}$ and $\widehat{NISL_{reassign}}$ are obtained by simulations with link reassignment techniques. As we can see, the simulation results are consistent with the theoretical analysis. In contrast, columns $\widehat{\delta_{fixed}}$、$\widehat{S_{fixed}}$ and $\widehat{NISL_{fixed}}$ present the snapshot information in satellite system using the fixed link assignment method[10]. Columns $\widehat{\delta_{equal}}$, $\widehat{S_{equal}}$ and $\widehat{NISL_{equal}}$ represent the snapshot information using the equal time interval snapshot partition techniques. To obtain the comparison accuracy, the time interval in equal time partition method is equal to the snapshot duration of link reassignment method, i.e. $\widehat{\delta_{equal}} = \widehat{\delta_{reassign}}$ = 274.31s/141.69s.

Compared to the results of fixed and equal time snapshot partition method, link reassignment method owns the improvements in the aspect of snapshot duration, number of snapshots, utilization ratio of inter-plane ISLs and end to end delay, especially for the Iridium system which contains odd number of satellites in each orbit plane:

- Constant and longer snapshot duration

As is shown in TABLE II. , the snapshot durations gained by link reassignment for both systems keep constant and is irrelevant to the polar border latitude. For Iridium system, the snapshot duration is the sum of the two successive snapshot durations of ordinary fixed link assignment method. For Teledesic system which contains even number of satellites in one plane, the two successive snapshot durations are uniformly divided to constant snapshot durations. As a result, constant and longer duration is more conducive to the performance stability of routing in satellite networks.

- Constant and half of the snapshot number

For the Iridium system, the number of snapshots obtained by link reassignment method (22) is half of the number obtained by the fixed link assignment (44). The reduction of snapshot number can decrease the times of routing table switches and gain more stable routing in satellite network. For Teledesic system, however, it stays unchanged also because of its even number of satellites in one plane, which originally has the same effect as the single event based snapshot partition.

- Higher utilization ratio of inter-plane ISLs

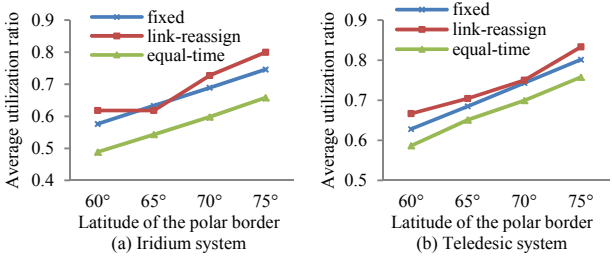

Fig. 6. Average utilization ratio of inter-plane ISLs

Finally, both satellite systems gain higher utilization ratio of inter-plane ISLs by using the link reassignment method. With the values in TABLE II. , we compared three methods' utilization ratio (U) of inter-plane ISLs which can be calculated as follows,

$$U = \frac{\sum_{i=1}^{S} NISL_i * \delta_i}{(N-1)*M*T} \quad (9)$$

where the denominator denotes the available time of all inter-plane ISLs without any ISL on-off changes, and the numerator denotes the available time of current inter-plane ISLs under a specific snapshot partition method. As is shown in Fig. 6, only when the polar border latitude is 65° in Iridium system, the ratio of the link reassignment method is slightly lower than the fixed link assignment method. The decrease is mainly due to the snapshot with large number of inter-plane ISLs (35) takes up relative large snapshot duration (261.91s) in fixed method than the link reassignment method with lower constant ISL number (34) for a constant duration (274.31s). In conclusion, although single event based snapshot partition can obtain improvement on the snapshot duration, snapshot number and ISLs utilization ratio, the uniform and constant snapshots in return lose the chance to gain the maximal number of ISLs.

- Lower average end to end delay

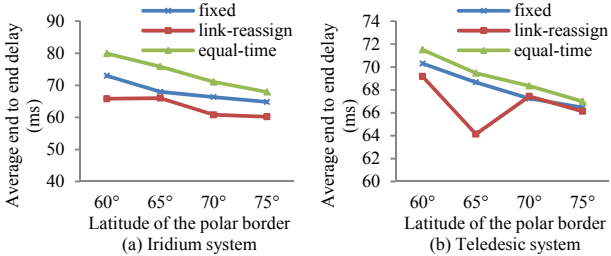

Fig. 7. Average end to end delay

Simulation results are shown in Fig. 7. The end to end delay is obtained from Beijing to London with packets sending in every 60s lasting for 24 hours. In most polar border latitude settings (except 70° in Teledesic system which is very close), average end to end delay of link reassignment is always lower than the other two methods. This is because the establishment of horizontal ISLs can provide shorter end to end paths. The snapshot duration of equal time method is selected as the duration of the link reassignment method, i.e. $\widehat{\delta_{equal}} = \widehat{\delta_{reassign}} = 274.31s/141.69s$. Because some non-persistent ISLs are deleted in equal time partition method, the end to end delay is highest among the three methods.

## V. CONCLUSIONS

Snapshot distribution is one of the key factors to the routing performance in satellite networks. In this paper, by using the single snapshot partition event (i.e. ISL break down event, ISL establishment event), we propose a novel link reassignment based snapshot partition method for better snapshot distribution. Simulation results show that our link reassignment method can keep the snapshot duration and ISL number constant, largely enhancing the routing stability. Meanwhile, this method has a shorter end to end delay compared to other two methods. Especially, even better snapshot distribution performance can be obtained for the Iridium system. Potentially, the Iridium-next system could be benefited from our method.

ACKNOWLEDGMENT

We are grateful to ICC 2015's reviewers whose comments will help us improve this paper.